# Natural Encoding of Information through Interacting Impulses

Vladimir S. Lerner, USA

*Abstract*—**Natural encoding is interactive process between a source of a natural energy and its receptor where the energy erases entropy of the interacting process impulses encoding the process impulses in information bits.**

**Standard unit of information Bit generates natural process through discrete (yes-no) interactions, including interactive macro impulses in classical physics and elementary micro impulses in quantum interactions.**

**Multiple interactions generate random process of interacting impulses with step-down and step-up actions.**

**The elementary interactions of the random step-up and step-down actions measures (yes-no) probability events according to Kolmogorov 1-0 law.**

**In a natural interactive process, each impulse step-down action cuts the process correlation of prior interactive events. That cuts the process entropy, defined by its probability.**

**The cutting entropy, at maximal probability, by cost of cutting energy memorizes information hidden in the correlation.**
**Memory "freezes" the information impulse as Bit.**

**The interactive impulse reveals information as phenomenon of interactions.**

**Each interval of interaction, holding needed quantity and quality of energy, naturally extracts its Bit and encodes information in the interacting physical state (by the impulse step-up action) carrying the energy cost. .**

**The interacting actions curve impulse whose curvature creates asymmetry and allows encoding logical qubits in memorized bit. The asymmetry lowers cost of energy needed for memorizing bit.**

**The impulse sequential natural encoding merges memory with the time of memorizing information and compensates the cutting cost by running time intervals of encoding in information process.**

**The information process, preserving the invariant impulse cutting information, binds the impulse' inner reversible microprocesses in the multiple impulses of the macroprocess' information irreversible dynamics.**

**The encoding process describes an information path functional integrating the receptor geometrical cellular information structure in a rotating code helix that composes the sequencing cells bits.**

**This natural encoding satisfies Landauer principle and compensates for the cost of Maxwell Demon.**

**The quantity and quality energy of specific interaction, necessary for natural encoding, limits the universal code length and density.**

**The applied entropy functional and information path integral measure the natural interacting process on its micro and macro trajectories.**

**The applied Jarzynski equation describes stochastic thermodynamics within the impulse microprocess (quantum) and connects this equation with information invariant measure in the encoding.**

**The thermodynamics of curved impulses on the rotating microprocess trajectories describe forming physical micro units encoding qubits, bits.**

**The increasing correlation connections of interacting events speeds their time course, while decreasing correlation slowing their time course. The time correlated entangled states stay coupled forever.**

**The relative moving information receptor possesses both relative encoding and attractive curved gravitation, emerging in the microprocess in addition to relative time and space.**

**The found encoding conditions and theoretical results validate the computer simulation, multiple experiments in coding genetic information, experimental encoding by spiking neurons, others.**

## INTRODUCTION. ABOUT NATURAL ORIGIN OF INFORMATION AND ITS NATURAL ENCODING

For many years of controversies, notion of information has not scientifically conclusive definition, neither implicit origin. Actually, information emerges naturally.

Information Bit holds yes-no $\downarrow\uparrow$ inter-action which naturally defines the elementary information unit.

The fundamental interactions build the structure of Universe. Numerous multilevel interspecies interactions self-organize biosystems.

Human interactions unify these and many others, often studied as the transmission of Information.

*Physical reality is only interactions, identified or not yet.*

Thus, information is *universal physical substance–a phenomenon of interaction*s which not only originates information but transfers it sequentially.

The sequential natural transfer information evaluates the level of its order-disorder or symmetry-asymmetry, which determines the energy quantity and quality of specific interaction. The sequence identifies a chain of natural sources and receptors of transferring information, and the information encoding which limits the code length.

In such sequence of natural transformation, each following reaction becomes an elementary virtual observer of the preceding action.

The sequence of the interactions builds a natural observing process where each elementary interactive impulse $\downarrow\uparrow$ is doing such observation.

Natural encoding, proceeding in the sequence of interacting processes, is analogous to creating new elements in chemical chain reactions. The elements enclosing components of the reaction "memorize" the interactive yes-no result similar to encoding.

Since the Bit emerges in interacting impulse, which is encoded naturally, the question is how to adjoin both natural emergence information and its encoding, which Nature had produced building DNA.

The introduced information formalism of both natural emergence of information and its encoding, also shows their advantage over non-natural encoding.

In the considered approach [I, others arxiv.org.], the sequence of interactions models Markov process which itself is product of many natural physical process, whose observation starts from Brownian motion.

Therefore, the approach unifies the natural origin of information with its natural model of observation.

The question is still how an observer, interacting with environment, naturally encodes each emerging impulse, and integrates multiple of them in the observer code?

Encoding observers in a universal code enables communicating with other information observers and recognizing their encoded natural information. It becomes possible when a creating Information Observer integrates the observing impulse' chain of information sources and receptors, sequentially encoding the impulse natural information. Such artificially build coding structure could understand other already encoded Information Observer' coding structure including its emerging genetic code.
In these different artificial observers, the natural interaction creates information usable for human communication, mutual acceptance, and understanding of human being.
Each impulse encoding connects their multiple chains.
The multiple interactions become random, and their process observes Markov diffusion with transitional probabilities and entropy measure.
Connection the entropy measure with the natural interactive information is subject of this approach.

In the study of thermodynamics [II], notion entropy was introduces, as a measure of irreversibility in a cyclic thermodynamic process. The generalized measure of this entropy [III] lead to the mathematically related measure in theory of communication [IV], called the entropy measure of transferring information.
Sill meaning of physical entropy is illusive.

The approach shows [V, others] that each impulse of interaction measures a *time interval* along the observing (interacting) process and *equivalent amount of information* generated by the interactions during this time.
Since interactions are everywhere, as well as its related time, the interactions become a measure of this time, and vice versa.
The *time obtains a physical nature of information.*
Intensive interactions curve its time, and the curved time initiates an interactive space, which *also obtains physical nature and measure.*
The natural encoding itself is an interactive process whose information and arrow of time initiate the interactions that carry energy for a following encoding.
This paper consolidates these concepts.

REFERENCES TO INTRODUCTION

## I. THE PRINCIPLE OF NATURAL ENCODING INTERACTIVE INFORMATION IMPULSES IN AN OBSERVER'S INFORMATION STRUCTURE

Regular computations either run out of time or run out of memory [1] and cannot concurrently encode information during natural interaction with environment.

Such interactions model random step-up and step-down impulse actions, which measures (Yes-No) probability events according to Kolmogorov 1-0 Law.

Markov diffusion process models a random process collecting multiple inter-active impulses of natural process. The diffusion encloses a set of the interactive events holding correlations, which cover a hidden entropy.

The sequence of the probabilistic 0-1 (No-Yes) impulses, acting on the Markov diffusion process, initiates Bayes probabilities that virtually self-observe the Markov process.

These objective probabilities link the Kolmogorov Law's axiomatic probabilities with the Bayesian probabilities, bringing discrete Bayes probabilities into the observing process.

The discrete probabilities virtually cut the Markov correlations which keep an entropy measure. Sequentially cutting the entropy of correlation decreases the uncertainty of Markov diffusion and increases the probabilities of the observing process.

During the random impulse interactions emerges principle of max-min entropy, where the probability maximum brings the entropy minimum.

Each impulse $\downarrow\uparrow$ action $\downarrow$ cuts the Markov maximal probability, opening a path to certainty, while the following interaction $\uparrow$ cutting the maximal entropy carries an equivalent unit of Information. Maximal probability of interactive impulses $\downarrow\uparrow$ enclose the certainty of real observation, bringing the actual interactive energy to the real interacting cut, which *converts* the maximal entropy to the Information unit.

The Information Bit uncovers informative events of the observing process hidden under the cutting correlations.

Multiple interacting Bits self-organize the Information process, whose integration creates the Information Observer.

The observing Bayes probabilities of multiple interacting events $\downarrow\uparrow$ hold inner probabilistic min-max logic of decreasing entropy. With converting entropy to information a certain information logic arises.

The observing process entropy on its random trajectories integrates Entropy Functional (EF) [2], formulas (1-2,5) in Sec.III. The EF also measures the process correlations with their time intervals equivalent to the entropy (6-8, 13-14).

Imposing the minimax principle on the EF leads to related Variation Problem (EP), which solves the equations of Interactive Integrated Information Dynamics, III (22-26). The dynamic trajectories allow prediction the proceeding interacting natural process.

The growing entropy density in the observing process curves an emerging ½ time units of the standard Kronicker's impulse time interval and initiates its displacement on two space units, being a counterpart to the curved time. (More details in Sec.II and [16].)

But the EP preserves invariant impulse entropy measure independently of size scale and curvature during the dynamics of the observing natural interactive movements.

The impulse correlation of the curving cut is orthogonal [46].

Within the observing process, the growing Bayes' a posteriori probability along neighbor impulses may merge, generating interactive jump on such impulse border. The merge meets action with reaction, superimposing cause and effect and their probabilities. It could cover unpredictable events within the merge.

A starting jumping action $\uparrow$ interacting with opposite $\downarrow$ action of the bordered impulses initiates the impulse inner process called a microprocess.

The opposite rotating transitive actions generate conjugating entropies of the microprocess (22-22a). (Details are in [4], [16]).

The rotation movement of finite action $\uparrow$ with the preceding jump $\downarrow$ settles a temporal transitional impulse $\uparrow\downarrow$ (inside the main interacting impulse), which launches correlation, rising entanglement of the conjugated entropies.

The entanglement is finalizing at angle of rotation π/2 forming double a space entropy unit with a volume within the transitional impulse.

Thus, *the entanglement, that initiates the curve time, starts before the space is formed and ends with beginning the space.*

The transitional impulse holds transient actions $\uparrow\downarrow(\pm)$ opposite to the primary impulse $\downarrow\uparrow(\mp)$ with the conjugated entanglement, involved, for example, in left and rights rotations ( $\mp$ ).

The transitional impulse, interacting with the opposite correlated entanglements $\mp$ , reverses it on $\pm$ .

Such logical operation sets a logical of entropy unit.

Since the hidden entropy' impulse is virtual, transition action within this impulse is also virtual, and its interaction with the forming correlating entanglement is reversible.

The curving interactive movement virtually erases each previous directional rotating the entangle entropy units creating asymmetrical logic of the entangle qubit.

At beginning of such impulse starts rotating cut of step-down action $\downarrow$ that injects an external energy which is finite, transits along the cutting entropies, and ends the curving impulses with erasure the cutoff entropy, finalizing the natural encoding process with increasing the impulse' curvature.

This erasure emits minimal energy $e_l$ of quanta

$$\varepsilon_o = \hbar\omega_o, e_l = \varepsilon_o[\exp(\varepsilon_o / k_B\theta) - 1]^{-1}$$

which lowers the energy quality compared with the injected energy ( $\hbar$ is Plank's constant, $\omega_o$ is frequency, $k_B$ is Boltzmann constant, and $\theta$ is absolute temperature).

This emission absorbs the transitional impulse asymmetrical logic, which logically memorizes the entangle qubit by making asymmetrical mirror copy of the information qubit.

Such operations perform function of logical Maxwell Demon [4].

The entangle logic is memorized temporary until the rotating step-up action $\uparrow$ , ending the transitional impulse, moves to transfer the entangle entropy qubit to the ending step-up action that kills the entropy and finally memorizes the joint entangle qubits in the impulse ending state as the information Bit.

(The temporal memorizing allows retains both entropy qubit with reversible logic and irreversible information qubit with asymmetrical logic.)

The killing is the irreversible erasure encoding the Bit that satisfies the Landauer principle [7] which requires energy from an external irreversible process compensating for the real cost of Maxwell Demon.

Logical energy $e_l$ , in a classical-macroprocess' limit (at $\hbar \to 0$ ), is transformed to real energy of the elementary Bit: $e_r = k_B\theta$ acting as the microprocess inner thermodynamics.

The microprocess, creating the entangled qubit, has reversible time until it interacts with environment, which brings irreversibility with energy.

To exclude the interaction, creators of quits hold them in Black Boxes [43].

The invariant discrete information units, which being cut off from the EF, integrate an Information Path Functional (IPF) [3-5], formulas (15-16). The IPF evaluates total *time* interval (17) of the information process starting with cutting entropy up to the natural encoding.

Conclusively, the impulse step-down cut $\downarrow$ , extracting the Bit hidden position, erases it at a cost of the cutting real time intervals along the evolving process, which encloses the entropy and energy of the natural interactive process.

The impulse step-up' $\uparrow$ stopping state at the end of the final impulse's time interval, delivering this energy, memorizes and encodes the information Bit of that impulse.

From (15-17) it seen that the impulse encoding merges its memory with the time of encoding, which minimizes that time.

The IPF integrates the encoded Bits in information macroprocess (23), while EF predicts next cutting correlation (9).

The Bit sequence memorizes their correlation' connection which keeps the Bits cooperation in a double and triple during the interactive observation [4].

The IPF extremals analytically describe these operations revealing that the triplets are the optimal macrounits of the extreme minimax information macrodynamic process.

The time-space information macrodynamic process cooperates the natural encoding Bits in the integral information geometrical structure of the *triplet code.*

The multi-dimensional interactive process initiates a path from prior uncertainty-entropy to posterior information-certainty.

This interactive process evolves with growing memory in the impulse cutting process correlations.

Each correlated cut binds the impulse sequence and memorizes both their code logic and time following its length and space size. The code logic becomes a source of logical complexity [6].

Information complexity memorizes that logic.

The process resolution–conversion to each posterior Bit in the observing process automatically encodes the cooperative logical and information geometrical structure of an Information Observer [5]-a receptor of the natural encoding.

Since each impulse interaction through its elementary discrete inter-actions $\downarrow\uparrow$ defines the standard unit of information, the Bit, the invariant impulse' bit-code is universal.

The bits are able to originate in any natural process, generated through various "anatomies" [55, 56] of interacting impulses, which bring different code length, density, and curvature.

Each process' dimensional cut measures the IPF finite Feller kernel' [19] information [46], which, at infinite dimensions, approaches the EF measure, restricting maximal information of the Markov multi-dimensional diffusion process.

The variety of the impulse $\downarrow\uparrow$ physical interactions unites these impulses in the information code which EF-IPF integrates in the encoding information process, generally describing the diverse natural interactive processes via universal code logic.

In the natural encoding' interactive process, when the interactive energy erases entropy of the interacting process impulses, encoding the process impulses in multiple bits, it also encodes this universal logic in both the micro and macroprocess.

The interactive impulses' information process naturally connects its entropy cuts with encoding and memory. It progressively develops the encoding,

The energy quantity (power) and quality of specific interaction (evaluated by the level of its order-disorder or symmetry-asymmetry) limits the universal code length by a final Bit's information density.

These principles substantiate the theoretical basis for different natural encoding in universal code of specific length, density, which is feasible at reasonable limitations.

## II. VERIFICATION AND DISCUSSION

According to Landauer principle [7], any logically irreversible manipulation with information, such as encoding, leads to erasure the information in a dissipative irreversible process.

Erasure of information Bit requires spending entropy

$S = k_B \ln 2.$

Maxwell Demon (MD) [8] associates the entropy of Landauer's principle with the acquisition of information Bit equals to the entropy' inevitable cost for the erasure.

Erasure of Bit requires at least work

$$W_B = k_B \theta \ln 2$$

at absolute temperature $\theta$, which theoretically limits the finite energy' resources or the time of performing such operation.

Bennett [9] found that any computation (encoding) can be performed using only reversible steps, which in principle requires no dissipation and no power. However, specific reversible computer needs to reproduce the map of inputs to outputs, erasing everything else that requires the energy cost.

Information cannot be copied with perfect accuracy according to no-cloning principle [10]; observing the copy disturbs the original states of the system, creating the errors requiring erasure.

Bell [11, 12] shows that information can be encoded in nonlocal correlations between the different parts of a physical system, which have interacted and then separated.

In Shor algorithm [1], the properties of the correlations between the "input register" and "output register" of computer functions require huge memory to store. This nonlocal information is hard to decode, the time of $n$ operations grows faster than any power of $\ln(n)$.

Practically such computations either run out of time or run out of memory.

The information defines the memorized entropy (uncertainty) cutting from the correlation hidden in interactions whose impulses process the encoding.

That definition connects the impulse entropy at the localities of the physical origin of information with its encoding, memory and energy cost.

The physical impulse interactive process enables progressively develop the encoding information operations, memorizing in the process current real time carrying equivalent entropy at the cost of the cutting time energy.

Applying the Jarzynski equality (JE) of irreversible thermodynamic transition [13] to conversion energy in information, and using results of its experimental verification [14], lead to the JE in form

$$e^{\Delta F/k_B\theta} - < e^{W/k_B\theta} >= \gamma.$$

Here $\Delta F$ is increment of free energy needed to produce energy $W, \gamma$ is parameter of the verification, which defines sum of probabilities $P_\Sigma \Rightarrow \gamma$ that inverse trajectory is observed in the experiments.

At $\gamma = 1$, the JE satisfies exactly:

$$e^{\Delta F/k_B\theta} - < e^{W/k_B\theta} >= 1, P_\Sigma = 1.$$

A thermodynamic process, satisfying the JE for all its states sequence, evolves irreversibly from the microprocess.

The quantity of information $I_{\delta t}$ at a curved cut has equivalent of free energy $\Delta F = \Delta F_{\delta t}$ to produce energy $W_t$ during a fixed time $\delta_t$, satisfying relation

$$\Delta F_{\delta t} = k_B \theta I_{\delta t}.$$

Average exponential energy $< \exp W / k_b \theta >$, collected during natural interactions at interval $\delta T_t$, measures the EF integral increment $\Delta S_{\delta t}$:

$$< \exp W / k_b \theta >= \exp \Delta S_{\delta t}.$$

Erasing that entropy by the natural source' high quality energy brings a non-random information $I_{\delta t}$ measures of the curved impulse during $\delta_t < \delta T_t$.

Influx of impulse energy $\Delta F_{\delta t}$ at $\delta_t$ enables converting entropy $\Delta S_{\delta t}$ to equivalent information $I_{\delta t}$.

Substituting both in the JE:

$$\exp I_{\delta t} - \exp \Delta S_{\delta t} = 1,$$

brings the equivalence of JE in both formulas for transition the entropy to information at probability $P_\Sigma = 1$.

*It confirms our initial statement about connection the entropy and information and condition of their conversion.*

By emerging real time on cutting intervals $\delta_t$ a certain logic with $I_{\delta t}$ appears with probability $P_\Sigma = 1$, which brings entropy $\Delta S_{\delta t} = 0$ (relates to $\ln(P_\Sigma = 1) \Rightarrow \Delta S_{\delta t} = 0$).

Substitution that in the previous equation leads to $\exp I_{\delta t} - 1 = 1, I_{\delta t} = \ln 2.$

Or, the equivalence of JE in both formulas for the transition to information requires invariant $I_{\delta t} = 1 Bit,$ the unit of information per impulse to compensate for the Maxwell Demon energy at the time interval of transition.

Thus, to satisfy the MD, the information produced by each impulse time interval should be invariant, holding constant the unit (Bit, Nat) in $I_{\delta t}$.

It confirms that the impulse minimax extreme principle (EP) satisfies the JE for impulse information transition, or vice versa. Each impulse time interval enables encoding invariant unit of information.

Or, the EP follows from the JE in the physical process whose interactive time interval is an equivalent of the impulse information cutting from the correlation carrying the above energy.

The cutting correlation's time intervals hold the information equivalent of this energy, and any real time interval of interaction brings the entropy equivalent of energy $\Delta F_{\delta t}$ which compensates for the MD while producing information during the interaction.

In interactive random process, whose sequence of cuts satisfy the EP, each impulse encodes the cutting correlation, and all information of the process cutoff correlations encodes the information process fulfilling the minimax law which is independent on size and curvature of any impulse.

Moreover, with probability $P_\Sigma = 1$, the inverse conjugated trajectories of the interacting impulses in the microprocess are part of the natural interacting process. That exactly satisfy the JE initial conditions [13-14] according to [16, 4].

The evolving microprocess starts with the merging impulse probability which satisfies relational entropy $S^*_{\mp a} = 2$ of the inverse states.

The 0-1 entropy units (potential bit) of the microprocess impulse connect the impulse inner correlation, while 01-0-1 entropy entity (a potential qubit) binds the microprocess' entanglement.

Multiple microprocesses, which the observation generates, hold statistical thermodynamic process where the JE automatically measure energy of these impulses discrete units.

*The relations, measuring energy within the curved impulse microprocess (quantum), connecting the JE with encoding this process' information measure at the cutting correlation, where applied the JE for the first time in [45].*

*The random interactions on the path to generation information naturally average the impulse microprocess' dissipative work in the JE thermodynamics.*

*The curved impulse' thermodynamics on the rotating microprocess trajectories describe forming physical micro units encoding qubits, bits.*

This approach distinct from other JE applications by *averaging the work in the JE during the evolving natural encoding*, while others need multiple experiments and specific procedure of averaging their results.

Such an impulse natural encoding merges memory with the time of memorizing information and compensates the cutting cost by running time intervals of encoding.

The information process' last cutting impulse encodes the process total information integrated in its IPF.

The encoding process, preserving the invariant cutting information, connects its multiple bits or qubits in the invariant irreversible thermodynamics where each discrete information unit' energy measures the JE. Thus, the JE allows measuring energy of the entropy or information unit in statistical thermodynamic microprocess and encoding thermodynamic macroprocess.

The invariant cutting information process holds the invariant irreversible thermodynamics measured through piece-vise Hamiltonian in (24-24a) [3] and diffusion-kinetic equations (23). Here, at $L_t \ge 2b_t$ kinetic flow transfers to diffusion $2b_t$, at $L_t \le 2b_t$ the diffusion flow transfers to kinetics, where the transformation applies on a small $\varepsilon$-localities of the bordered impulse.

These conditions are found in [54] where, at kinetic transition, the Hamiltonian includes increments of chemical potentials of interacting physical-chemical entities.

Hence, the JE, with connection to (23), describes increments of temperature, entropy, energy, diffusion, physical-chemical components in the variety of thermodynamic processes within interacting impulses and their cooperative macrodynamics, which hold equivalent information description.

Each impulse (Fig.1b) step-down action has negative curvature $-K_{e1}$ corresponding attraction, the step-up reaction has positive curvature $+K_{e3}$ corresponding repulsion, the middle part of the impulse, having negative curvature $-K_{eo}$, transfers the attracting entropy of the statistical thermodynamics between these parts.

Thus, the JE applies to the curved impulse thermodynamics on the rotating microprocess trajectories forming physical micro unit encoding qubit, bit. [53].

When an external process interacts with the natural impulse, it injects energy capturing the entropy of impulse' ending step-up action.

This inter-action generates next impulse' step-down reaction, modeling 0-1 bit (Fig.1a,b).

The considered opposite curved interaction provides a time–space difference (an asymmetrical barrier) between 0 and 1 actions, necessary for creating information Bit.

The interactive impulse' natural step-down ending state memorizes Bit, where the interacting process supplies Landauer's energy with maximal probability closed to 1.

The step-up action of natural process curvature $+K_{e3}$ encloses potential entropy $e_o = 0.01847 Nat$, which carries entropy $\ln 2$ of the impulse total entropy $1Nat$ and may transit it to interacting (external process).

The interacting step-down part of the external process impulse' invariant entropy $1Nat$ has potential entropy $1 - \ln 2 = e_1$. Actually, this impulse' step-down opposite interacting action brings entropy $-0.25Nat$ with anti-symmetric impact $-0.025Nat$ which carries the impulse wide entropy $\sim -0.05Nat$ [16] while the total $\sim -0.3Nat$ is equivalent to $-e_1$.

Thus, during the impulse interaction, the initial energy-entropy $W_o = k_B \theta_o e_o$ changes to $W_1 = -k_B \theta_1 e_1$. Since the interacting parts of the impulses have opposite: positive and negative curvatures accordingly; the first one repulses, the second attracts the impulse energies.

The external process needs minimal entropy $e_{10} = \ln 2$ for erasing the Bit, which corresponds Landauer's energy

$$W_B = k_B \theta \ln 2 .$$

If the interactive external process accepts this Bit by memorizing (through erasure), it should deliver the Landauer energy compensating the difference of these energies-entropy: $W_o - W_1 = W_B$ in balance form

$$k_B \theta_o e_o + k_B \theta_1 e_1 = k_B \theta \ln 2.$$

Assuming the interactive process supplies the energy $W_B$ at moment $t_1$ of appearance of the interacting Bit, it leads to equality $k_B \theta_1 (t_1) = k_B \theta (t_1)$.

That brings the balance to forms

$$k_B \theta_o 0.01847 + k_B \theta (1 - \ln 2) = k_B \theta \ln 2,$$ or

$$\theta_o / \theta = (2 \ln 2 - 1) / 0.01847 = 20.91469199.$$

Natural impulse with maximal entropy density $e_{do} = 1 / 0.01847 = 54.14185$, interacting with external curved impulse, transfers minimal entropy density $e_{d1} = \ln 2 / 0.01847 = 37.52827182$.

Ratio of these densities

$k_d = e_{do} / e_{d1} = 1.44269041$ equals to $k_d = 1 / \ln 2$.

The opposite curved anti-symmetric interaction, with ratio $(2 \ln 2 - 1) / \ln 2 \cong 0.5573$,

decreases ratio of the above temperatures on amount of $\ln 2 / 0.0187 - (2 \ln 2 - 1) / 0.01847 = 16.61357983.$

Here the impulse's interacting curvature, enclosing this entropy density, lowers the initial energy and the related balanced temperatures in the above amount. From that follow

*Conditions creating a bit in interacting curved impulse:*

1. The opposite curving impulses in the interactive transition require keeping entropy ratio 1/ln2.
2. The interacting process should possess the Landauer energy by the moment ending the interaction.
3. The interacting impulse should hold invariant measure [1] of entropy 1 Nat whose the topological metric preserves the impulse curvatures. •

The last follows from the impulse' max-min mini-max law under its stepdown-stepup actions, which generate invariant [1] Nat's time-space measure' topological metric π (1/2circle) preserving opposite curvatures.

Results [15] prove that physical process, holding invariant entropy measure for each phase space volume ($v_{eo} \cong 1.242$ per process dimension in [4]), characterized by above topological invariant, satisfies Second Thermodynamic Law.

Energy $W_B$ that delivers an external process will erase the entropy for both attracting and repulsive movements, covering energy of the both movements, which are ending at the impulse stopping states.

The erased impulse' total cutoff entropy is memorizes as equivalent information, encoding the impulse Bit in the impulse ending state.

The natural step-up action captures its entropy, while moving along the action positive curvature and transiting to interacting step-down action' negative curvature, gains the impulse logic, and by overcoming entropy-information gap [4, 16] acquires the equal information that compensates for the movements logical cost.

Thus the attractive logic of an invariant impulse, converting its entropy to information within the impulse, performs function of *logical Demon Maxwell* in the microprocess.

*Topological transitivity at the curving interactions*

The impulse of the natural process holds its $1Nat$ transitive entropy until its ending curved part interacts, creating information bit during the interaction.

Theoretically, when the cutting maximum of entropy reaches a minimum at the end of the impulse, the external interaction occurs, converting the entropy to information by getting the energy from the interactive process.

The invariant' topological transitivity has a duplication point (transitive base) where one dense-form changes to its conjugated form during orthogonal transition at hitting time. During the transition, the invariant holds its measure (Fig.1b) preserving its total energy, while the densities of these energies are changing.

The *orthogonal* transitive base separates the primary dense form and conjugate dense form of the topological transition.

At the transition turning moment, a jump of the time curvature switches to a space curvature (Fig.1a) with potential rising space waves [16] in the microprocess . •

As a distinction from traditional MD, which uses an energy difference in temperature form [8]**,** this approach reveals a MD through naturally created difference of their curvatures.

Forming transitional impulse with the entangled qubits leads to possibility memorizing them as a quantum bit.

That requires first to provide asymmetry of the entangled qubit, which starts with the anti-symmetric impact by the main impulse step-down action $\downarrow$ interacting with opposite action $\uparrow$ of starting transitional impulse. This primary anti-symmetric impact $-0.025 \times 2 = -0.05Nat$ curve both main and transitional impulses with curvature $K_{e1} \simeq -0.995037$, enclosing $0.025Nat$, while the starting step-up action of the transitional impulse generates curvature $K_{e2} \simeq +0.993362$ enclosing $e_o = 0.01847Nat.$

Difference $(0.025\text{-}0.01847)Nat$ estimates entropy measuring total asymmetry of main impulse $0.00653Nat = S_{as}$ .

The entangled qubit in the transitional impulse evaluates entropy volume 0.0636 Nat, which defines the spending entropy on transfer the minimal entangle phase volume $v_{eo} \cong 1.242$ to the entropy-information gap [16], while primary impulse impact brings minimal entropy $0.05Nat$ starting the entangled curved correlation.

Thus, the correlated curved entanglement can memorize $(0.05 - 0.0656)Nat$ in the equivalent information of two qubits.

The middle part of the main impulse generates curvature $K_{e2} \simeq +0.993362$ which encloses entropy $0.02895Nat.$

Difference $0.02895 - 0.025 = 0.00395Nat$ adds asymmetry to the starting transitional entropy, while $0.02895 - 0.01845 = 0.0105Nat$ estimates the difference between the final asymmetry of the main impulse and ended asymmetry of the transitional impulse.

With starting entropy of the curved transitional impulse $0.05Nat$, the ending entropy of the transitional impulse' asymmetry estimates

$$0.0653\text{-}0.0105\text{-}0.00395\text{=}0.05085Nat .$$

Memorizing this asymmetry needs compensation with a source of equivalent energy. It could be supplied by opposite actions of the transitional step-down $\downarrow$ and main step-up interacting action $\uparrow$ ending transitional impulse.

That action will create the needed curvature at the end of the main impulse, adding $0.0653 - 0.05085\text{=}0.01445, 0.01445\text{-}0.0105\text{=}0.00395Nat$ to entropy of transitional impulse curvature sum $0.05085$.

Another part will bring the difference of entropy' curvature $0.02895 - 0.01845 = 0.0105$ with total 0.0653.

Thus, $0.05085 Nat = s_{asv}$ is entropy of asymmetry of entropy volume $s_{ev} = 0.0636$ of transitional impulse, whereas $0.0653 Nat = S_{as}$ is entropy of asymmetry of the main impulse.

This asymmetry generates the same entangled entropy volume that step-action of the main impulse transfer for interacting with external impulse.

Thus, $s_{asv}$ is the information "Demon cost" for the entangled correlation, which the curvature of the transitional impulse encloses.

Using the asymmetrical curvature of transitional impulse that holds the entangle volume, enclosing the entangle correlation, instead of direct evaluation this correlation, allows memorizing information of two qubits in impulse measure 1 Nat.

That evaluation is closed to [26], obtained differently and confirmed experimentally.

When the posteriori probability is closed to reality, the impulse positive curvature of step-up action, interacting with the merging impulse' negative curvatures of step-down action, transits a real interactive energy, which the opposite asymmetrical curvatures actions enfolds.

During the curved interaction, this primary virtual asymmetry compensates for the asymmetrical curvature of a real external impulse, and that real asymmetry is memorized through the erasure by the supplied external Landauer's energy.

The ending action of external impulse creates classical bit with probability

$$P_k = \exp -(0.0636^2) = 0.99596321.$$

Since the entanglement in the transitional impulse creates entropy volume $0.0636$, the potential memorizing pair of qubits has the same probability.

Therefore, both memorizing classical bit and pair of qubits occur in probabilistic process with high probability but less than 1, so it happens and completes not always.

The question is how to memorize entropy enclosed in the correlated entanglement, which naturally holds this entropy and therefore has the same probability?

If transitional impulse, created during interaction, has such high probability, then its curvature holds the needed asymmetry, and it should be preserved for multiple encoding with the identified difference of the locations of both entangled qubits.

Information, as the memorized qubits, can be produced through interaction, which generates the qubits within a material-devices (a conductor-transmitter) that preserve curvature of the transitional impulse in a Black Box, by analogy with [43].

At such invariant interaction, the multiple connected conductors memorize the qubits' code.

The needed memory of the transitional curved impulse encloses entropy $0.05085 Nat$.

*The time intervals of the curved interaction*

When the natural space action curves the internal interactive part, the joint interactive time-space curved action measures its interactive impact.

If the interaction at moment $t_o$ creates internal curvature $K_{e1} \simeq -0.995037$ enclosing $-0.025 Nat$ by moment $t_{o1} = 0.01845 Nat$, then interacting time-space interval measures the difference of these intervals

$$|t_{o1}| - t_o \triangleq 0.0250 - 0.01847 = 0.00653 Nat.$$

For that case, the internal curved inter-action attracts the energy of natural interactive action.

If No part of the interacting impulse emerges at $t_o$ and Yes part arises by $t_1$, then the invariant interacting impulse will spend $1 - \ln 2 + \ln 2 = 1$ Nat on creation a bit ($\ln 2 Nat$).

If inter-action of the natural process on the internal process delivers energy $W_B$ by moment $t_1$, this energy will erase the bit and memorize it according to the balance relations above.

The interacting impulse spends~1 Nat on creating and memorizing bit $\ln 2$ while holding free information $(1 - \ln 2) \cong 0.3 Nat$.

The curved topology of interacting impulses decreases the needed energy ratio according to the above balance relation. Thus, time interval $t_o - t_1$ creates the bit and performs the MD function.

Multiple interactions generate *a code* of the interacting process at the following *conditions*:

1. Each impulse holds an invariant probability–entropy measure which the natural Bit code also satisfies.
2. The impulse interactive process, which delivers such code, must is a part of a real physical process that keeps this invariant entropy-energy measure equivalent to metric π.

That process memorizes Bit and creates information process of multiple encoded Bits, building process' information dynamic structure.

For example, a water, cooling natural drops of hot oils in the found ratio of temperatures, enables spending a part of the water energy' chemical components to encode other chemical structures. Or the water kinetic energy will carry the accepting multiple drops' bits as an arising information dynamic flow.

3. Building the multiple Bits code requires increasing the impulse information density in three times with each following impulse acting on the interacting process [16]. Such physical process generating the code should supply the needed energy for free information of three bits which sequentially attracts each other.
Each interactive impulse, produced a Bit, should hold three impulses measure π, i.e. frequency of interactive impulse should be f=1/3 π=~0.1061.•

The interval 3π provides opportunity to join three bits' impulses in a triplet, as elementary macrounit, and combats a noise and redundancies from both natural and internal processes.

Ratio of the impulse density Nat/Bit=1.44 to its average curvature $K_{e2}$ is equivalent to a relative information density mass: $M^m = 1.44 / K_{e2}$.
For the asymmetrical impulse' curvature $K_{e2} = 0.993362$, $M^m = 1.452335645$.

The opposite curved interaction lowers potential energy, compared to other interactions for generating both a Bit, and the information mass.

The multiple curving interactions create topological bits code which sequentially forms moving spiral structure on Figs. 2-3.

The curving interaction dynamically encodes Bits in natural process developing information structure Figs. 2-4 of the interacting information process.

*How to find an invariant energy measure, which each bit encloses starting the MD?*
Since the bit's requires minimal energy $W_B = k_B \theta \ln 2$, it's possible, following above relation $W_B = F_{\delta t} = (\mathrm{k}_b\,\theta)[1]$, to find such temperature $\theta_1^o$ that equals the inverse value of $k_B$.
If the interacting process carries this temperature, then its minimal energy holds $W_B^o = \ln 2$ at $\theta_1^o = 1 / k_B$, which equals the invariant time-space Nat measure of physical bit , or its entropy logic measure.

Let us evaluate $\theta_1^o$ at $k_B = 8617 \times 10^{-5}\,\mathrm{eV/K}$ and Kelvin temperature
$K = 20 / 293 = 0.0682259386^{oC/K}$ equivalent to $20^{oC}$. Then $\theta_1^o = 588.19 \times 10^5 /\mathrm{eV}$.
If we assume that this primary natural energy brings the eV amount equivalent to quanta of light $e_q = 1240\mathrm{eVnm}$, $1\mathrm{nm}=10^{-9} m$, then we come to $\theta_1^o = 588.19 \times 10^5 \times 1.240 \times 10^3 / e_q \times 10^{-9} m \cong 72.9356^{oC/m} / e_q$.
Or each quant should bring temperature' density $\theta_1^o = 72.9356^{oC/m}$, which is reasonably real.
With this $\theta_o^o$, the interacting impulse will bring energy $W_B^o = \ln 2$ to create its bit.
Following the balance relation, the external process at this $\theta_o^o$ should have temperature $\theta_o^o = 20.91469199\theta_1^o = 1525,42^{oC/m}$ brought by a quant. This energy holds an invariant impulse measure $|1|_M = 1Nat$ with metric π, or each such impulse has entropy density $1Nat / \pi$.
The bit of the interacting impulse has minimal density energy equivalent to $\ln 2 / \pi = 0.22$ at temperature $\theta_1^o$.

The impulse cutting action of the growing density curves an emerging ½ time units of the impulse time' interval measure $|1|_M$ while a following rotating curved time-jump initiates a displacement within the impulse opposite rotating Yes-No actions [4].

That originates a space shift, quantified by the curved time. While the impulse holds invariant Kolmogorov's probability (1 or 0) for two space units (as a counterpart to the curved time, Fig.1).
The Fig.1a shows forming a space unit during the curved time-jump according to relation
$$2\pi h[l] / 4 = 1 / 2 p[\tau]$$
for space coordinate/time ratio $h / p$, which leads to ratio of the measures for the time and space units:
$$[\tau] / [l] = \pi / 2$$
with elementary space curvature equals to inverse radius
$$K_s = h[l]^{-1}.$$
Thus, the jump' squeezed time interval originates both curvature and space coordinate.
When the two space units replace the curved ½ time units within the same impulses, such transitional time-space impulse preserves measure $|2 \times 1/2|_M = |1|_M$ of the initial time impulse.
The considered rotating *transitional* impulse allows encoding the microprocess' qubit on a middle spot of the curved impulse.
The curving impulse ↓↑ gets form (Fig.1b) whose curvature holds transitional information and complexity.
The implementation of the minimax principle leads to sequential assembling of a manifold of the process' elementary information units in binary units (doublets) and then in the triplet macrounits.

The free information of interacting bits attracts them through a resonance of coherent frequencies in the asymmetrical curved interaction.

The EF-IPF integrates the progressively curving impulse geometry in the rotating double space spiral trajectories, located on a conic surface of a three-dimensional equation of the extremals. Each spiral segment encodes the bits of the information triplet macrounits enclosing the impulses information microprocesses. The spiral encodes multiple impulses cooperating in macroprocess [16].

Manifold of the extremal segments, cooperating in the triplets' optimal structures, forms an information network (IN) with a hierarchy of its nodes (Fig. 2), where the IN accumulated information is conserved in invariant form.

The information transformed from each IN's previous triplet to the following one (in the IN hierarchy) has an increasing value, because each following triplet encapsulates and encloses the total information from all previous triplets.

The sequential cooperation of the IN triplet's nodes creates the information structure which condenses the total node produced information in the IN ending node (which includes its free information).

The node unique time-space location within the IN hierarchy determines the value of information encapsulated into this node.

A sequence of the successively enclosed triplet-nodes, represented by discrete control logic, creates the IN code with the three digits from each triple segments and a forth from the control that binds the segments.

The code serves as a virtual communication language and an algorithm of minimal program to design the IN.

The optimal IN code has a double spiral (helix) triplet geometrical structure (DSS) simulated on Fig.3, which is sequentially enclosed in the IN final node, allowing the reconstruction of both the IN dynamics and topology.

The IN upper node hierarchy' measure, through its free information, automatically requests for higher information values from the observing process cutting correlations. Information density of the cutting correlations' space curvature generates information force of an adaptive feedback, which attaches each requested information to the IN [5].

The IN information geometry holds the node's binding functions and the asymmetry of triplet's structures.

In the DSS information geometry, these binding functions are encoded, adapting the requested external information to encode.

The Observer self-builds the IN information space–time networks, which hierarchically enfolds multiple observing information triplets encoding the Observer logical structure in triplet code (Fig.4). Hence, the information of observing process moves and self–organizes the information geometrical structure creating the Information Observer.

The natural encoding information during different interactive processes of an observer with environment (at limitations [4]) produces the self-arising information Observer.

The DSS specifics depends on the structure of the EF integrant' drift and diffusion, in (1) and (2), during the interactions.

Minimal entropy ln2 classifies quality of energy (from the high-quality light energy to the low-quality energy of heat dissipation).

Since various interactions carry energy different quantity and quality, the impulse universal logical code enfolds the particular power of these quantities and qualities. When each impulse cutoff captures the specific needed energy, it quality determines the current cutting entropy, while the energy quantity defines the information density of encoding Bit.

Therefore, each interactive process encodes the specific information code sequence ending with maximal Bit's information density, which limits the code length.

The encoding energy quality determines the observing process entropy (19) through the probability when encoding of the impulse *information* starts.

This probability $P_m \cong 0.985507502$ is limited by minimal uncertainty measure $h_\alpha^o = 1/137$ -the physical structural parameter of energy [17], which includes the Plank constant's equivalent of energy. It also counts a sub-Plank spot being uncertain during a Bayes probability of the interacting impulses (20). (Specifically such probability of a No-action possibly covers a microprocess under the sub-spot [47],[16]).

After entropy volume of the growing probability increases to overcome the uncertain measure of the interacting process, the entropy reaches the edge of certainty-reality with an ability of increasing the probability up to 1 and revealing the process information.

Markov process, modeling interactive natural process, whose cutting sequence satisfies the EP, may naturally reveal information process from these interactions, which the IPF encodes in its Feller kernel.

The $n$-dimensional process cutting off generates a finite information measure, integrated in the IPF whose information approaches the EF measure at $n \to \infty$. That restricts maximal information of the Markov diffusion process and the ability of encoding, which limits maximal information density of the code unit.

Comments. Number $M$ of the equal probable possibilities determines Hartley's quantity of information $H = \ln M$, which for the impulse $M = 2$ holds

$H = \ln 2 Nat.$

The impulse information measured in Bits holds

$I = 1/\ln 2 \ln M = 1 bit.$

The correlation cutting by the impulse brings information $0.75 Nat$ from which $\delta S_u \cong 0.0568 Nat$ delivers the impulse cut.

Minimal physical time interval limits light time interval

$\delta t_\tau \cong 1.33 \times 10^{-15}$ sec

defined by the light wavelength

$\delta l_m \cong 4 \times 10^{-7} m.$

That estimates maximal impulse information density by

$I_{ko_k} \cong \ln 2 / 1.33 \times 10^{-15} \cong 5.2116 \times 10^{+15} Nat/s.$

The IPF integral information evaluates maximal density enclosing all information in the finite impulse time interval, which is the impulse cutting time instant, delivering the correlation's hidden information $\ln 2$.

All integrated information enfolds the Feller kernel whose time and energy evaluate results [18].

The EF integral (1) on trajectories of Markov diffusion process measures both process entropy and time interval (11), (12) of the process correlations in (8). Each cutting the EF freezes the probability of the interacting events of the process.

The EF presents a potential informational path functional of the Markov process until the applied impulse, carrying the cutoff contributions, transforms it to the IPF.

The multi-dimensional delta-action on the EF (1) multi-dimensional integrant-additive functional (2) allows analytical solution the impulse encoding and its representation by Furies series.

Each IPF dimensional cut measures the finite Feller kernel information. The IPF concentrates integrates all interactive information in a final finite Kronicker's impulse-discrete analog of Dirac delta-function with values 0 and 1 (20), which structure Bit with the estimated maximal density.

Physical meaning of object, event ($O_f$) connects its probabilistic-information description trough $P(O_f) \to 1.\bullet$

*Relative information observer*

Let's analyze how impulse linear space speed $c_k$ on the IPF curved time-space trajectory–an extremal of the observing process affects the impulse with invariant geometrical measure $\pi$ which encloses its invariant information measure $|1|_M$.

Suppose impulse $\pi_k$ located on distance $l_k$ along the IPF trajectory is moving with linear speed $c_k$, and another invariant impulse $\pi_{k1}$ with the same invariant information measure, located on distance $l_{k1}$, is moving along this trajectory with linear speed $c_{k1} > c_k$ relative to impulse $\pi_k$. Each impulse speed is the ratio of invariant geometrical measure to the impulse location' time interval on the trajectory: $\tau_k$ and $\tau_{k1}$ accordingly, at $c_k = \pi_k / \tau_k, c_{k1} = \pi_{k1} / \tau_{k1}$, where at $c_{k1} > c_k$, $\pi_k = \pi_{k1} = \pi, \tau_{k1} < \tau_k$..

With growing time along the IPF trajectory, the impulse time intervals shortens $\tau_{k1} < \tau_k$. It leads to persisting increase of information densities: $D_k = |1|_M / \tau_k, D_{k1} = |1|_M / \tau_{k1}$ of the impulse with invariant information measure $|1|_M$: $D_{k1} > D_k$.

Distance between these impulses on the trajectory is $\Delta l_{k,k1} = l_{k1} - l_k = (c_{k1} - c_k)/(\tau_{k1} - \tau_k),$

$\Delta l_{k,k1} = \pi(\tau_k - \tau_{k1})^2 / \tau_{k1}\tau_k, \tau_{k1} < \tau_k.$

Since impulse preserves invariant measure $\pi$, this distance limites condition $(\tau_k - \tau_{k1})^2 / \tau_{k1}\tau_k \geq 1$.

Minimal distance at this condition satisfies $(\tau_k - \tau_{k1})^2 / \tau_{k1}\tau_k = 1, (\tau_k - \tau_{k1})^2 = \tau_{k1}\tau_k$, which after solution quadratic Eq. $\tau_k^2 - 3\tau_{k1}\tau_k + \tau_{k-1}^2 = 0$ leads to minimal admissible time ratio $\tau_{k1} / \tau_k = (3 - \sqrt{5})/2, \tau_{k1} / \tau_k \cong 0.38$, or related maximal admissible time ratio $\tau_{k1} / \tau_k = (3 + \sqrt{5})/2, \tau_{k1} / \tau_k \cong 2.36$.

In both cases, at $\Delta l_{k,k1} = \pi$, the decreasing time ratio satisfies only the first solution. Moreover, this ratio *limits minimal distance* of observation each invariant impulse $\pi$. With growing speed, at decreasing this ratio, the observing impulse does not preserve its invariant measure. Therefore, this ratio limits both distance of observation each invariant impulse and all observing path along the trajectory.

The path of observation of the invariant impulse from location $(l_{k1}, \tau_{k1})$ is minimal on the trajectory.

These nearest $\pi_k$, $\pi_{k1}$ impulses hold third one $\Delta l_{k,k1} = \pi = \pi_{k,k1}$ between them. These three *identify a triple located* on each $k$ dimension of $n$-dimensional observing process $n = 1, 2, 3, ..., k, ...$ Such triples possess above features.

At the minimal time interval, more bits concentrate in the impulse time unit which grows the speed of the natural encoding. The growing information density increases the impulse curvature along the trajectory.

The raising curvature and impulse speed of natural encoding enfolds the growing impulse information mass.

The information observer defines the encoded information impulses.

The information observer, moving with maximal speed, encodes maximal information with maximal speed of natural encoding, combined with growing curvature, density, information mass, and *shortening path of observation.*

The IPF integrates the density in observer' geometrical structure (Fig.4). The structure' rotating speed grows with increasing the linear speed, and shortening time intervals of the invariant impulse information measure.

The growing density increases number of the enclosed events in both time and space intervals. Both initially emerge in the micro process in addition to relative time and space.

Hence, the observing minimal impulse time interval automatically increases its linear speed, curvature, density, speed of encoding, shortens information mass, which links to growing attraction in an information gravitation on the reduce path of observation.

The relative moving information observer possesses relative encoding and attractive curved gravitation.

Formulas $X_\delta = -1/4 r_x^{-1/2}$ and $r_{icM} = \pi \times K_{eo}$ [4] connect the *week information force*, arising from correlations, with the topological curvature:
$X_{\delta cM} = -1/4(\pi \,|\, K_{eo} \,|)^{-1/2}$.

At $r_{icM} = r_x$, above relation connects a *strong information force* $X_t$ in (23) with the macroprocess curvature, which links to a potential observing space gravitation.

Topological curvature turns to information curvature, and leads to connection of information force $X_t$ with the information curvature [16].

The observer relative time encloses information and the curvature (gravitation). That connects the information force with information mass and complexity according to formulas (25-26).

Hence, the information approach connects the considered relative times of moving observers, the observing process' time, the microprocess time, and certain-real time of information observers.

The inner and external time courses in information observer are mutual connected.

The observing sequence of time intervals initiates its logic which connects the observer cognition with the observer coding logic and intelligence [52].

For the information observer, the increasing time holds more encoding information.

*Finally, all these follow from the fundamental phenomena of interaction.*

The discussed approach adds information *versions* to Einstein's theory of relativity applied to moving interactive information observer and emerging its inner time course.

<u>Comments.</u>

Each abstract axiomatic Kolmogorov's probability predicts probability measurement in the experiment whose probability distribution, tested by the events occurrences' relative frequencies, satisfy symmetry condition of the equal probable events [19].

In the probability field, sequence of random events $\omega_\eta$, collected on independent series, forms Markov chain [19] with multi-dimensional probability distribution.

The initial Kolmogorov probability distribution of probability field and the following EF (1) represent all $n$-dimensional Markovian model of the interacting process.

Probabilities of each process dimension are local for each its random ensemble being a part of whole process ensemble. All process dimensions may start instantly but with different local probabilities associated with local random frequencies.

Local probabilities of each random ensemble are symmetrical, and Markov process describes Kolmogorov equations for direct and inverse transitional probabilities.

The correlated values of $n$-dimensional impulse' entropies emerge as process' probabilistic nonlocal logic created during the interactions, which processing and encoding multiple nonlocal qubits and bit [16].

The microprocess emerges inside random process described by sub-markov diffusion process [48] when the impulse actions bring negative entropy measure $S^*_{\mp a} = -2$ with relative probability $p_{a\pm} = \exp(-2) = 0.1353$.

The interactive jump initiates the microprocess at reaching minimal relative time difference at the displacement edge [16], where the microprocess satisfies only multiplicative probabilities as quantum process.

The microprocess time in Quantum Mechanics is reversible until interaction–measurement affects quantum wave function.

Real (physical) arrow of time arises in natural macroprocesses which average the multiple microprocesses with their reversible local time intervals making a temporal "hole" in the arrow of a macroscopic time.

Natural arrow of time ascends along the multiple interactions and persists by the process growing correlations.

Both virtual and information observers hold own time arrow: the virtual - symmetric, temporal, the information - asymmetric physical, which memorize the natural encoding observer's information.

Whereas the total time direction holds, each a non-locality of the quantum microprocess provides reversible time-space holes, while processing the irreversible time-space impulses admits localities which acquire the energy of the random field.

Since particular interaction accesses only a part of entire random field, the process interaction' both particular time interval and time arrow distinguish.

Thus, the time holds a discrete sequence of the impulses carrying entropy from which emerges a space in the sequence: interactions-correlations –time-space.

The real time processes memory encoding the process information with persistent logical causality for observer' time. •

## A. ILLUSTRATIONS

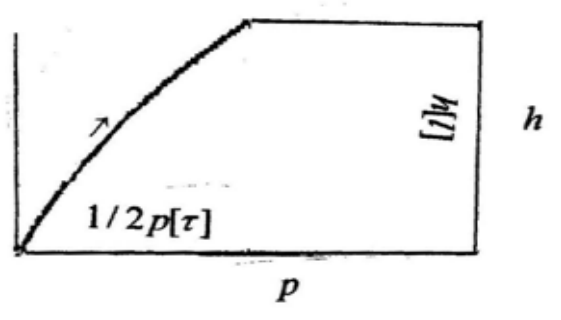

1(a)

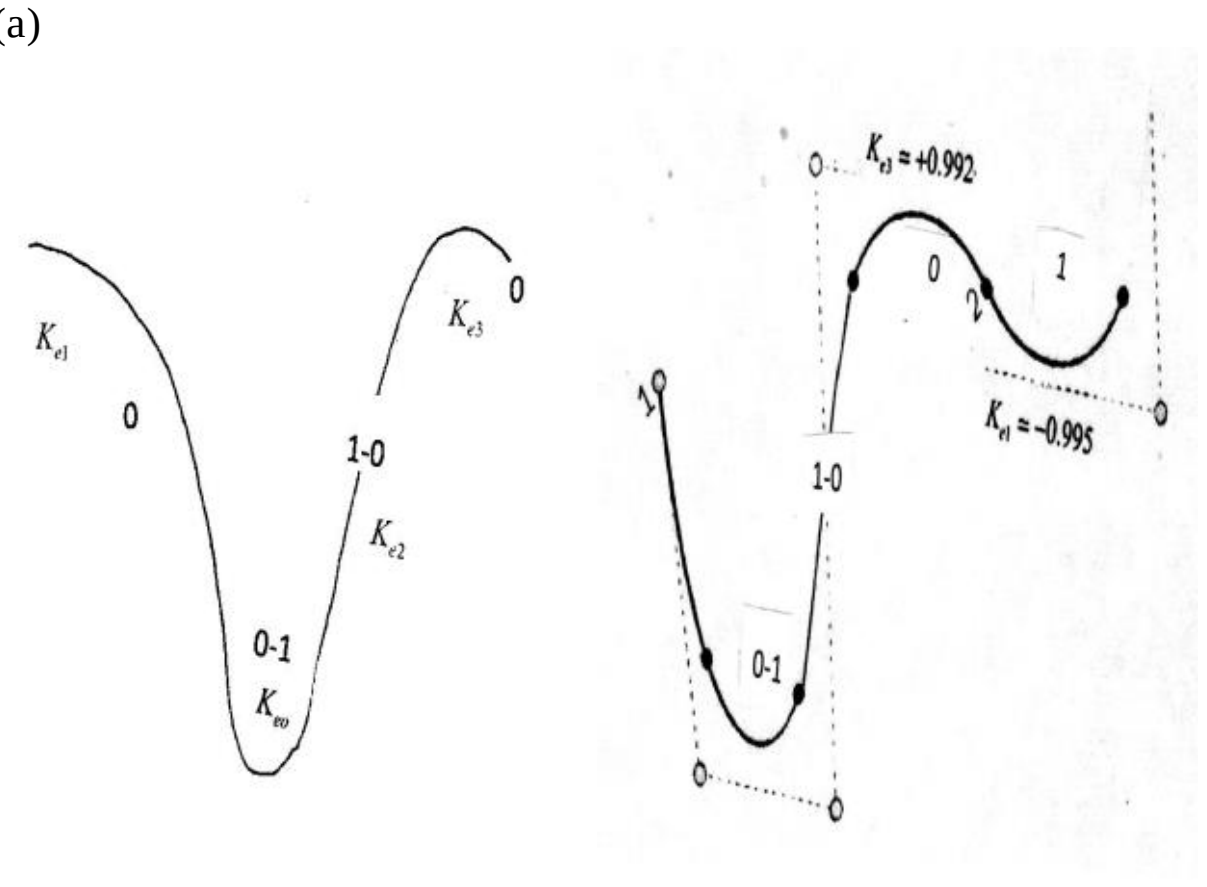

(b)

Fig. 1. (a).Illustration of origin the impulse space coordinate measure $h[l]$ at curving time coordinate measure $1/2p[\tau]$ in transitional movement.

(b).Curving impulse with curvature $K_{e1}$ of the impulse step-down part, curvature $K_{eo}$ of the cutting part, curvature $K_{e2}$ of impulse transferred part, and curvature $K_{e3}$ of the final part cutting all impulse entropy.

A virtual impulse (Fig.1b, left) starts step-down action with probability 0 of its potential cutting part; the impulse middle part has a transitional impulse with transitive logical 0-1; the step-up action changes it to 1-0 holding by the end interacting part 0, which, after the inter-active step-down cut, transforms the impulse entropy to information bit.

On Fig. 1b, right, the impulse in Fig. 1a, left, starting from instance 1 with probability 0, transits at instance 2 during interaction to the interacting impulse with negative curvature $-K_{e1}$ of this impulse step-down action, which is opposite to curvature $+K_{e3}$ of ending the step-up action $-K_{e1}$ is analogous to that at beginning the impulse Fig.1a).

## B. *COMPUTER SIMULATIONS*

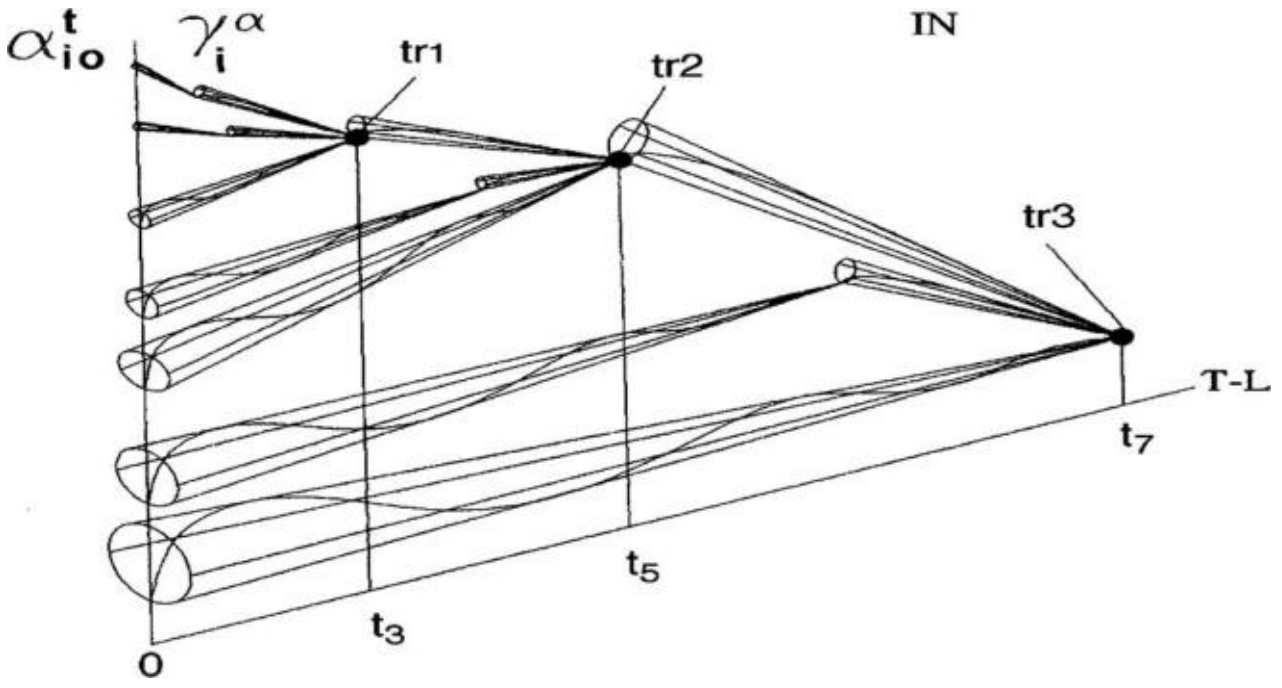

Fig.2. The IN information geometrical structure of hierarchy of the spiral space-time dynamics of triplet nodes (tr1, tr2, tr3,…); $\{\alpha_{io}\}$ is a ranged string of initial eigenvalues, cooperating in (t1, t2, t3) locations of T-L time–space, $\{\gamma_{io}\}$ is parameter measuring ratio of the IN nodes space-time locations.

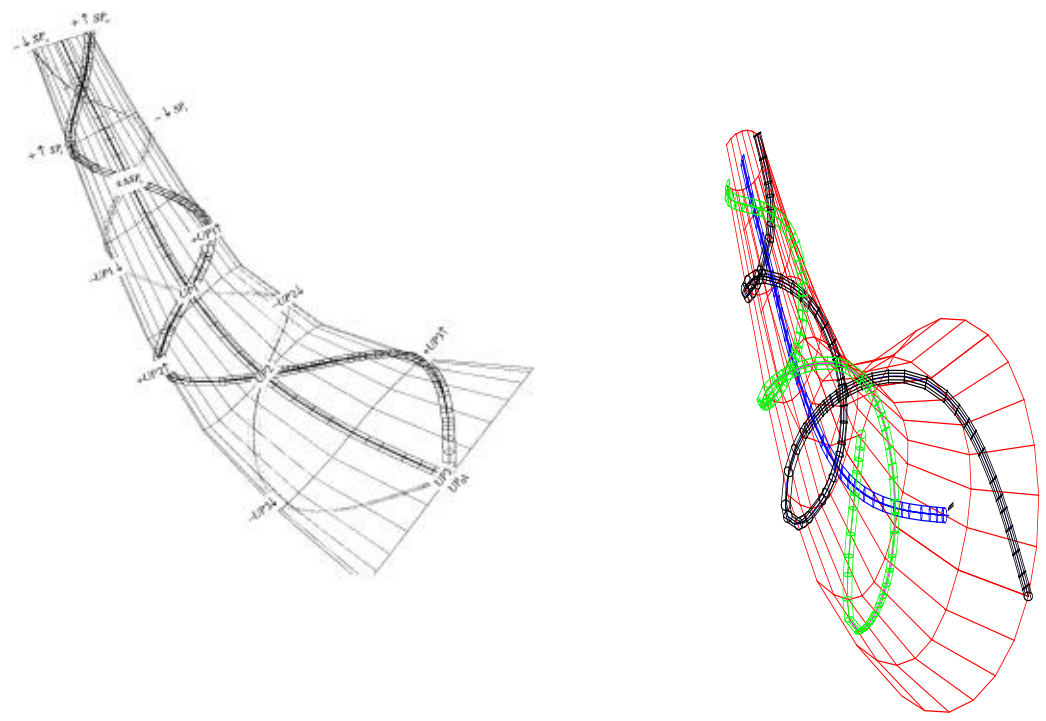

Fig.3. Time-space opposite directional-complimentary conjugated trajectories $+\uparrow SP_o$ and $-\downarrow SP_o$, forming the spirals located on conic surfaces. Trajectory on the spirals bridges $\pm\Delta SP_i$ binds the contributions of process information macro unit $\pm UP_i$ through the impulse joint No-Yes actions, which model a line of switching inter-actions (the middle line between the spirals).Two opposite space helixes and middle curve are on the right.

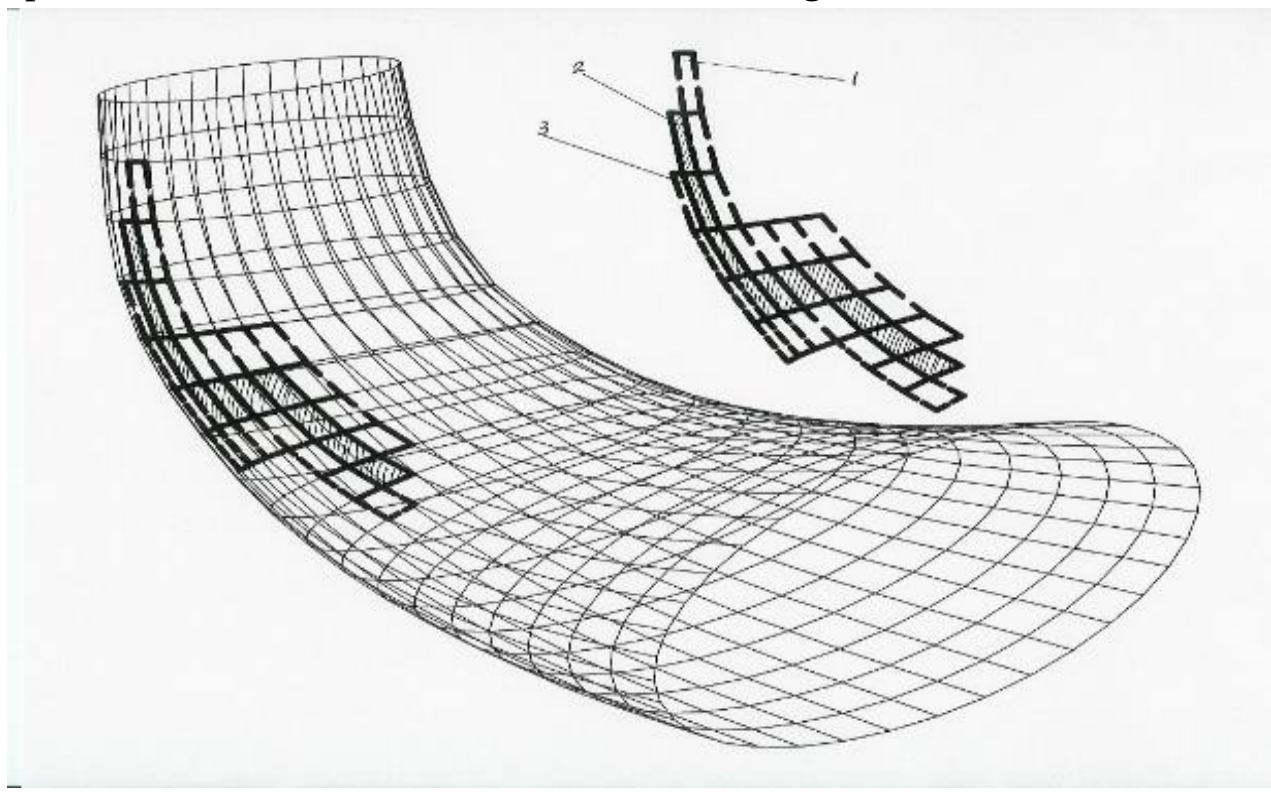

Fig. 4. Structure of the cellular geometry, formed by the cells of the DSS triplet's code, with a portion of the surface cells (1-2-3), modelling the space formation of Information Observer. This structure's geometry integrates information contributions simulating in Figs.2, 3.

## III. MATHEMATICAL FORMALISM OF INTERACTIVE INTEGRATED INFORMATION DYNAMICS (IIID)

IITD formalizes the description of information regularities of observing interactive process.

The IIID integrates the information path from randomness and uncertainty to information, thermodynamics, and the emerging intelligence of observer.

The integral measure of the observing process trajectories formalizes Entropy Functional (EF), which is expressed through the regular and stochastic components of Markov diffusion process $\tilde{x}_t$ [2]:

$$\Delta S[\tilde{x}_t]|_s^T = 1/2 E_{s,x}\{\int_s^T a^u(t,\tilde{x}_t)^T (2b(t,\tilde{x}_t))^{-1} a^u(t,\tilde{x}_t)dt\} =$$
$$\int_{\tilde{x}(t)\in B} -\ln[p(\omega)]P_{s,x}(d\omega) = -E_{s,x}[\ln p(\omega)], \quad (1)$$

where $a^u(t,\tilde{x}_t) = a(t,\tilde{x}_t,u_t)$ is drift function, depending on control $u_t$, and $b(t,\tilde{x}_t)$ is diffusion function determined by covariation function in Ito Eq. [20]. The EF integrant is the process additive functional [21]:

$$\varphi_s^T = 1/2\int_s^T a(t,\tilde{x})^T (2b(t,\tilde{x}))^{-1} a(t,\tilde{x}_t)dt + \int_s^T \sigma(t,\tilde{x})^{-1} a(t,\tilde{x})d\xi(t) \quad (2)$$

which describes transformation of the Markov processes' random time traversing the various sections of trajectory $\tilde{x}_t = \tilde{x}(t)$; $E_{s,x}$ is a conditional to the initial states $(s,x)$ mathematical expectation, taken along these trajectories.

Right side of (1) is the EF equivalent formula, expressed via probability density $p(\omega)$ of random events $\omega$, integrated with the probability measure $P_{s,x}(d\omega)$ along the process trajectories $\tilde{x}(t)\in B$, which are defined at set $B$.

The Markov diffusion is a physical process modeling a flow of particles exposed to random displacements at collisions with other particle and molecules.

In probability theory, Markov diffusion process is a solution to Ito' stochastic equations, which describes the Markov process transitional probabilities.

The EF integrates these probabilities along the trajectories of the Markov diffusion process allowing integrates both probabilistic and physical properties of this process.

Generally, a random process (as a continuous or discrete function $x(\omega,s)$ of random variable $\omega$ and time $s$), describes elementary changes of its probabilities from one distribution (a priori) $P_{s,x}^a(d\omega)$ to another distribution (a posteriori) $P_{s,x}^p(d\omega)$ in form of their transformation [22]:

$$p(\omega) = \frac{P_{s,x}^a(d\omega)}{P_{s,x}^p(d\omega)}. \quad (3)$$

Sequence of these probabilities ratios generalizes diverse forms of specific functional relations, represented by a series of different transformations.

The probability ratio in the form of natural logarithms:

$$-\ln p(\omega) = -\ln P_{s,x}^a(d\omega) - (-\ln P_{s,x}^p(d\omega)) = s_a - s_p = \Delta s_{ap} \quad (4)$$

describes the difference of a priory $s_a > 0$ and a posteriori $s_p > 0$ *random entropies*, which measure uncertainty, resulting from the transformation of probabilities for the process events, satisfying the entropy's additivity.

A change brings a certainty or information if its uncertainty $\Delta s_{ap}$ is removed by some equivalent entity call information $\Delta i_{ap}$ at $\Delta s_{ap} - \Delta i_{ap} = 0$.

Thus, information is delivered if $\Delta s_{ap} = \Delta i_{ap} > 0$, which requires $s_p < s_a$ and a positive logarithmic measure at $0 < p(\omega) < 1$.

Condition of zero information: $\Delta i_{ap} = 0$ describes a redundant change, transforming a priori probability to equal a posteriori probability; or this transformation is identical–informational undistinguished.

The removal uncertainty $s_a$ by $i_a : s_a - i_a = 0$ brings equivalent certainty or information $i_a$ about entropy $s_a$.

The logarithmic measure (4) of probabilities of the Markov diffusion process approximates the probability ratios for other random processes [23].

Mathematical expectation of random probabilities and entropies in (4):

$$E_{s,x}\{-\ln[p(\omega)]\} = E_{s,x}[\Delta s_{ap}] = \Delta S_{ap} \Rightarrow I_{ap} \neq 0 \quad (5)$$

determines mean entropy $\Delta S_{ap}$ as equivalent of nonrandom information $I_{ap}$ of a random source.

Being averaged by the source events, through a probability of multiple random variables-states, or by the source processes (through probabilities in (1)- depending on what is considered a process, or an event), both (5) and (1) include Shannon's formula for relative entropy-information of the states (events).

For a continuous random variables, (5) brings also an equivalent of Kullback–Leibler's (KL) divergence measure [24], expressed by a nonsymmetrical logarithmic distance between the related entropies in (5),(4).

The KL measure is connected to both Shannon's conditional information and Bayesian inference of testing a priori hypothesis by the observation of a priori-a posteriori probability distributions.

Markov diffusion process, with its statistical interconnections of states, represents the most adequate formal model of the information process, where functional (1) includes the Bayes probability link to the process trajectories with given drift and diffusion.

The probability link concurrently updates and integrates each a priori to following a posteriori probability along the process.

The EF integrant in (1) is partially observable measuring only covariation function on the process' trajectories.

For a single-dimensional EF (1) with drift function $a = c\tilde{x}(t)$, at given nonrandom function $c = c(t)$ and diffusion $\sigma = \sigma(t)$, the EF acquires form conditional to process "white noise" $\varsigma_t$ which has the same diffusion as the initial process, but the zero drift:

$$\varsigma_t = \int_s^t \sigma(v, \xi_v) d\xi_v.$$

Such conditional EF holds formula

$$S[\tilde{x}_t / \varsigma_t] = 1/2 \int_s^T E[c^2(t)\tilde{x}^2(t)\sigma^{-2}(t)]dt =$$

$$1/2 \int_s^T [c^2(t)\sigma^{-2}(t)E_{s,x}[x^2(t)]dt = 1/2 \int_s^T c^2[2b(t)]^{-1} r_s dt, \quad (6)$$

where for the Markov diffusion process, the following relations connecting functions of dispersion and diffusion are true:

$$2b(t) = \sigma(t)^2 = dr/dt = \dot{r}_t, E_{s,x}[x^2(t)] = r_s. \quad (7)$$

These relations identify EF (1) on the observed Markov process $\tilde{x}_t = \tilde{x}(t)$ by measuring covariation (correlation) functions at applying positive function $c^2(t) = u(t)$; where (6) represents the EF functional expressed through a regular integral with the integrant:

$$A(s,t) = r_s[2b(t)]^{-1} = r_s \dot{r}_t^{-1}. \quad (8)$$

The EF integral (6) is described through $u(t)$ in form

$$S[\tilde{x}_t / \varsigma_t] = 1/2 \int_s^T u(t) r_s \dot{r}_t^{-1} dt. \quad (9)$$

The $n$-dimensional functional integrant in (8) follows from related $n$-dimensional covariations in (7) and dispersion matrix applying $n$-dimensional function $u(t)$.

Correlation function (7) on small time interval $o(s)$:

$$r(s) = \int_s^{s+o(s)} 2b(t)dt = 2b(s)o(s) \quad (10)$$

leads to function (8) at this interval

$$A(s,s) = [2b(s)]^{-1} 2b(s)o(s) = o(s),$$

and to integrant (8):

$$A(s,t) = o(s)b(s)/b(t) = o(s,t), \quad (11)$$

which brings integral (9) to form

$$S[\tilde{x}_t / \varsigma_t] = 1/2 \int_s^T u(t) o(s,t) dt. \quad (12)$$

If function $u(t)$ cuts off the diffusion process on time interval $\delta_o = o(s,t)$, it cuts correlation function (10) of function (11). That brings entropy of cutting correlation to form (12).

Both (8) and (11) are forms of integrant for functional (2).

The impulse $\delta$-cutoff of action $u(t)$ evaluates the quantity of information which the functional EF conceals, when the correlations between the non-cut process states where bound.

The cutoff leads to dissolving the correlation between the process cut-off states, losing the functional connections at these states.

Applying delta-function $c^2(t, \tau_k) = \delta u_t(t - \tau_k)$ to integral (12) determines the cutting entropy functions:

$$\Delta S[\tilde{x}_t / \varsigma_t]\Big|_{t=\tau_k^{-o}}^{t=\tau_k^{+o}} = \begin{cases} 0, t < \tau_k^{-o} \\ 1/4 o(\tau_k^{-o}), t = \tau_k^{-o} \\ 1/4 o(\tau_k^{+o}), t = \tau_k^{+o} \\ 1/2 o(\tau_k), t = \tau_k, \tau_k^{-o} < \tau_k < \tau_k^{+o} \end{cases} \quad (13)$$

at $\tau_k^{-o} < \tau_k < \tau_k^{+o}$.

The cutoff brings direct measure of (1):

its maximum

$$S[\tilde{x}_t / \varsigma_t]_{t=\tau_k} = 1/2 o(\tau_k) = 1/2 Nats$$

on the impulse left border's interval $o(\tau_k)$:

$$S[\tilde{x}_t / \varsigma_t]_{t=\tau_k^{-o}} = 1/4 o(\tau_k^{-o}) Nats$$

and its minimum

$$S[\tilde{x}_t / \varsigma_t]_{t=\tau_k^{+o}} = 1/4 o(\tau_k^{+o}) Nats$$

-transferring to the right border interval.

Summary of the impulse intervals:

$$\sum_{t=\tau_k^{-o}}^{t=\tau_k^{+o}} \Delta S[\tilde{x}_t / \varsigma_t]_t = 1/4 o(\tau_k^{-o}) + 1/2 o(\tau_k) + 1/4 o(\tau_k^{+o}) = o_k. \quad (14)$$

evaluates the invariant $Nat$ fraction of the EF cutoff contribution on interval $o_k$, which the interval encloses.

In such impulse, represented through opposite No-Yes (0-1) actions, each No action carries the cutting impulse part with a maximum of cutting entropy, while Yes action, following the impulse cutting part, gains the maximal cutting entropy reduction.

The impulse $\delta_o = o(s,t)$ cut of correlation $r(s)$ at moments $s$ maximizes this entropy cutting part.

The correlation maximal jump at following current moment $t$ dissolves mutual correlation $r(s,t) \to 0$. That maximizes its derivation in (7), while minimizing part $\dot{r}_t^{-1}$ of the entropy integrant (8).

That validates max-min principle of relational entropy between the impulse points $(s,t)$ transferring probabilities (3).

The max-min variation principle implies the invariance of functionals (9), (12) under $u(t)$.

Sequential cuts transform the entropy contributions from each maximum though minimum to the next maximal information contributions, where each next maximum decreases at the following cutoff moments.

Each $\delta$-cutoff at these points loses the amount of 0.5 Nats minimizing current integral (12).

The equations of max-min variation principle for the EF describes extremal trajectories of information process, which the optimal EF integrates.

Information path functional (IPF) unites the discrete information cutoff contributions $\Delta I[\tilde{x}_t / \varsigma_t]_{\delta_k}$ taking along $n$ dimensional Markov process:

$$I[\tilde{x}_t / \varsigma_t]|_s^{t \to T} = \lim_{k=n \to \infty} \sum_{k=1}^{k=n} \Delta I[\tilde{x}_t / \varsigma_t]_{\delta_k} \to S[\tilde{x}_t / \varsigma_t]. \quad (15)$$

The IPF along the cutting time correlations on optimal trajectory $x_t$ in the integral upper limit is

$$I[\tilde{x}_t / \varsigma_t]_{x_t} = -1/8 \int_s^T Tr[(r_s \dot{r}_t^{-1}] dt = -1/8 Tr[\ln r(\mathrm{T}) / r(s)]. \quad (16)$$

The EF and IPF in (15) are equalizing in the limit:

$$\lim_{t \to T} \sum_{t=s}^{t=T} \Delta S[\tilde{x}_t / \varsigma_t]_t \to (T - s) \quad (17)$$

through the process time interval, which follows from the EF definition [2] by additive functional (2).

The IPF encoded Bits have maximal theoretical admissible probability density (3) concentrating the IPF in Feller kernel during the above minimal time and related space intervals.

Conditional Kolmogorov probability

$$P(A_i / B_k) = [P(A_i) P(B_k / A_i)] / P(B_k)$$

after substituting an average probability

$$P(B_k) = \sum_{i=1}^{n} P(B_k / A_i) P(A_i)$$

defines Bayes probability by averaging their finite sum or integrating [19].

For each $i,k$ random events $A_i, B_k$ along observing process, each conditional a priori probability $P(A_i / B_k)$ follows conditional a posteriori probability $P(B_k / A_{i+1})$.

Conditional entropy

$$S[A_i / B_k)] = E[-\ln P(A_i / B_k))] = [-\ln \sum_{i,k=1}^{n} P(A_i / B_k)] P(B_k) \quad (18)$$

averages the conditional Kolmogorov-Bayes probability for multiple events along the observing process.

Random current conditional entropy is

$$\tilde{\mathrm{S}}_{ik} = -\ln \mathrm{P}(A_i / B_k)\, \mathrm{P}(B_k). \quad (19)$$

The experimental probability measure predicts axiomatic Kolmogorov probability if the experiment satisfies condition of symmetry of the equal probable events in its axiomatic probability [19].

Conditional probability satisfies Kolmogorov's 1-0 law [19] for function $f(x)|\xi$ of $\xi, x$ infinite sequence of independent random variables:

$$P_\delta(f(x)|\xi) = \begin{cases} 1, & f(x)|\xi) \geq 0 \\ 0, & f(x)|\xi) < 0 \end{cases}. \quad (20)$$

This probability measure has applied for the impulse probing an observable random process, which holds opposite Yes-No probabilities – as the unit of probability impulse step-function [4].

The equations of the EF for the microprocess [16]:

$$\partial S(t^*) / \delta t^* = u_\pm^{t1} S(t^*),$$

$$u_\pm^{t1} = [u_+ = \uparrow_{\tau_k^{+o}} (j-1), u_- = \downarrow_{\tau_k^{+o}} (j+1)] \quad (21)$$

under inverse actions of function $u_\pm^{t1}$, starting the impulse opposite time $t_\pm^* = \pm\pi / 2t^i$ which measures space rotating angle relative to the impulse inner time $t^i$, determine the solutions-conjugated entropies $S_+(t_+^*)$, $S_-(t_-^*)$:

$$S_+(t_+^*) = [exp(-t_+^*)(\mathrm{Cos}(t_+^*) - jSin(t_+^*))]|,$$

$$S_-(t_-^*) = [exp(-t_+^*)(\mathrm{Cos}(-t_+^*) + jSin(-t_+^*))] \quad (22)$$

at

$$S_\pm(t_\pm^*) = 1/2 S_+(t_+^*) \times S_-(t_-^*) =$$

$$1/2[\exp(-2t_+^*)(\mathrm{Cos}^2(t_+^*) + Sin^2(t_+^*) - 2Sin^2(t_+^*))] = \quad (22a)$$

$$1/2[\exp(-2t_+^*)((+1 - 2(1/2 - Cos(2t_+^*))))]$$

$$= 1/2 \exp(-2t_+^*) Cos(2t_+^*)$$

The interactive entropy $S_\pm(t_\pm)$ minimizes the starting entanglement *which begins a space interval during reversible relative time interval of* $0.015625\pi$ part of the impulse invariant measure $\pi$.

By overcoming entropy-information gap starts the information Bit and Observer [ 4,16].

The EF-IPF minimax variation principle leads to Information Macrodynamic Eqs:

$$\partial I / \partial x_t = X_t, a_x = \dot{x}_t = I_f, I_f = L_t X_t, L_t = 2b_t \quad (23)$$

where $X_t$ is gradient of the IPF functional $I$ (15) on a macroprocess' trajectories $x_t$, $I_f$ is information flow defined through speed $\dot{x}_t$ of the macroprocess. The flow emerges from drift $a^u(t, \tilde{x}_t)$ being averaged by $a_x$ along all microprocesses, as well as the averaged diffusion $b_t \to b$ for the macroprocess force while. $L_t$ is kinetic matrix.

Information Hamiltonian $H$ :

$$-\frac{\partial \tilde{S}}{\partial t} = (a^u)^T X + b\frac{\partial X}{\partial x} + 1/2a^u(2b)^{-1}a^u = -\frac{\partial S}{\partial t} = H \quad (24)$$

determines the macroequations from the minimax variation principle.

Equations (23) are information form of the equations of irreversible thermodynamics, which the *information macrodynamics* generalizes.

The discretely changed information Hamiltonian divides irreversible dynamic trajectory on the partial reversible segments, predicting next emerging information unit.

The flows and forces determine the macroprocess Hamiltonian in form

$$H = X \times I . \quad (24a)$$

The EF-IPF connects the information curvature $K_\alpha^m$, the IN cooperative complexity $MC_m$, and effective cooperative complexity $MC_m^{\delta e}$, at forming the information structures, by Eqs:

$$K_\alpha^m = -M_{vm}^* = MC_m^{\delta e}, \quad (25)$$

where

$$MC_m^{\delta e} = 3\dot{H}_m^V MC_m \quad (26)$$

includes increment of Hamiltonian per volume $\dot{H}_m^V$ and the IN cooperative complexity $MC_m$.

*Single Equation (25) at (26) encloses all previous Eqs. unifying description of the approach formalism.*

The IPF is information form of Feynman path functional (FPF) in quantum mechanics, while EF integrates random entropy (uncertainty) and limits information (actual) contributions, including the time evolution in observing process.

The FPF is a quantum analog of action principle in physics, while EF expresses a probabilistic causality of the action principle, and the cutoff memorizes a certain information causality integrated in the IPF.

## IV. EXPERIMENTAL VERIFICATIONS AND APPLICATIONS

Natural increase of correlations demonstrates experimental results [25], [26].

Coding genetic information reveals multiple experiments in [27], [28].

Experimental coding by spiking neurons demonstrates [29].

Evolutions of the genetic code from a randomness reviews [30].

More such evidences are cited in [4, 16, 52].

That supports natural encoding through the cutting correlations and physically verifies reliability of natural encoding information process.

The impulse cut-off method was practically applied in different solidification processes with impulse controls' automatic system [31].

This method reveals some unidentified phenomena-such as compulsive appearance centers of crystallization-indicators of a generation information code, integrated in the IPF during the impulse metal extraction (withdrawing). (In such metallic alloys, the "up-hill diffusion, creating density gradients, is often observed" [32]).The frequency of the impulse withdrawing computes and regulates the designed automatic system to reach a maximum of the IPF information indicators. (Detail experimental data of the industrial implemented system are in [31] and [32]).

The automatic control regulator in the impulse frequency, cutting the process movement, was implemented for different superimposing electro-technological processes [33] which are interacting naturally. Examples of the method applications in communications, biological and cognitive systems, others are in [34], [35] and [36]. The developed computer program is in arXiv: 1303.0777.

Retinal Ganglion Cells are the Eyes discrete impulse receptors interacting with observations and generating information which transmission integrates [37].

Encoding through natural chemical reactions, connecting chemical molecules, are in [38].

Experiments [39] confirm encoding coherent qubits in spinning electron locked in attractive "hole spin". Other examples are quantum solar dots of semiconducting particles using for the information coding and retrieving images [40].

Experiments [49] confirms our results of tracing quantum particle by observing wave function probabilities.

In [57] the identified new triple photon structure is theoretically predicted in the author arXiv:1602.05129..

## V. INFORMATION SPECIFICS OF A NATURAL OBSERVER

NATURAL ORIGIN OF INFORMATION AND NATURAL ENCODING INFORMATION defines NATURAL OBSERVER

That includes

1. Natural emerging subjective bits
2. Natural creation information logic, cognition, and intelligence
3. Self-created consciousness.

These are components of IIID where

1. Natural correlated interactive Yes-No actions provide both Bits and Hidden information between them called Free Information, which the IPF integrates. Cooperative attraction, created by Free information, structures (at least 4.48 bits [16]) selects a minimal *subjective* observer-a triplet whose Free information enables adjoins other Bits in a logical structure during the cooperative attraction.

2. Equalizing speeds of the Free Information's attracting movement synchronize the frequencies of the triplet's logic in a resonance which assembles a logical loop. The multiple resonances of the logical loops assemble Information Network of hierarchical nodes of the nested triplet's logic.

Each nested coherent logical loop accepts only such units that each IN node logic recognizes, self-organizing a logical chain of *cognitive functions*, which are distributed along hierarchy of the assembling logical units: triplets, IN nested nodes, and the IN ending nodes.

The Observer cognition emerges as an evolving intentional ability to request, integrate, and predict the needed Observer Information that builds the Observer's growing networks. The hierarchy of loop logic is memorizing and then encoding after coordinated access of external energy to each local logical unit at every hierarchical level. The coordination performs the Observer own time clock with its time course, which commands the hierarchical switching.

The Observer hierarchical codes physically organizes the multiple IN, self-encoding their local codes in the Information structure of the *Observer intelligence.*

3. The self-creating observer's intelligence enables decoding each IN's node memorized information recognizing it in intercommunication associated with self-awareness. That precedes through the self-creation, initiating some observed bit or their cooperating triplets, whose free information of an attracting logic requests for cooperation with some logical information located in the specific IN node. Thus, the observer intelligence includes function of a self-awareness modeling consciousness.

The last result confirms philosophical [58] conclusion that term "consciousness" had no constant meaning, and should be abandoned, "it does stand for a function". …

*Studying the IIID, the author was not familiar with [58], coming to this result independently.

## VI. CONCLUSION

### A. SIGNIFICANCE OF FINDING

*1)* The standard unit of information Bit generates any natural process through discrete (yes-no) curved inter-actions, which include interactive macro-impulses in classical physics and elementary micro-impulses of quantum interactions.

*2)* The impulse natural inter-action cuts information hidden in correlation extracting each Bit hidden position, erases it at cost of cutting real time interval, and memorizes encoding information in the interacting physical state.

*3)* The asymmetrical difference curvature of natural interacting impulses lowers potential energy sequentially encoding and merging memory with time of encoding, which minimizes that time.

*4)* The interactive impulse reveals information to be a phenomenon of interaction.

*5)* The natural encoding includes transitional logical memory, which satisfies Landauer's principle, and compensates for the cost of Maxwell's Demon.

*6)* The energy of a specific interaction limits the universal code length and density.

*7)* The resolution–conversion of the impulse cutting entropy to information process and automatic encoding into a cooperative structure builds structure of Information Observer which satisfies the information form of a relative moving Observers.

### C. STEPS OF EMERGING THE INFORMATION OBSERVER

*1)* Reduction the process entropy under interacting impulses, observing by Bayesian probability, increases each posterior correlation; the growing correlations connect the observing process Bayes probabilities in probabilistic causality.

*2)* The impulse cutoff correlation sequentially converts the cutting entropy to information that memorizes the entropy logic in Bit which naturally encodes and participates in next entropy conversions as a primary Information Observer built without any a prior physical law.

*3)* The repeated interactions generate the information micro- and macro-processes, which governs the impulse natural minimax information law.

*4)* Elementary impulse interactive process creates reversible time and space intervals in emerging impulse microprocess with conjugated entangled entropy, curvature and logical complexity.Sequential interactive cuts integrate the cutting information in information macroprocess with irreversible time course.

*5)* The applied Jarzynski equation is measuring energy within the impulse microprocess (quantum) connecting this equation with encoding this process' information measure. This equation also measures energy of bits connected in macrodynamics process encoding the bits. The curved impulse thermodynamics on the rotating microprocess trajectories describe forming physical micro units encoding qubits, bits.

The JE also confirms our initial statement about connection the entropy and information and condition of their conversion.

*6)* The increasing correlation connections of interacting events speeds their time course, while decreasing correlation slowing their time course. Stronger correlation connection increases time of holding this connection, or time of coupling the correlated states together. The time correlated entangled states stay coupled forever.

*7)* The relative moving information observer possesses both relative encoding and attractive curved gravitation emerging in the microprocess in addition to relative time and space.

*8)* The encoding process unites the information path functional integrating observer geometrical cellular information structure which composes rotating helix of the sequencing cells Bits.

*9)* The memorized information binds reversible microprocess within impulse with irreversible information macroprocess of the multiple impulses. The transitive gap separates the micro- and macroprocess on an edge of reality.

*10*) The logical operations with the process information units integrate the discrete information hidden in the cutting correlations in information structure of the Observer. The process relational entropy conveys probabilistic causality with temporal memory of correlations, while the cutoff memorizes certain information causality in the objective probability observations. The observer Bit-Participator holds geometry and logic of its prehistory.

*11)* The self-organizing information triplet forms a macrounit of self-forming information time-space cooperative distributed network enables self-scaling, self-renovation, and adaptive self-organization.

*12)* The natural encoding impulse interacting process (virtual-imaginable up to real) creates a path from the process uncertainty toward certainty of real information Observer. The evolving Observer logic self-creates its conscience and intelligence [52].

*13*) The approach, starting with Kolmogorov's probabilities, creates the physical information micro and macro processes and the Observer without Physical particle theory.